# Primary Numbers Database for ATLAS Detector Description Parameters

A. Vaniachine, S. Eckmann, D. Malon
*Argonne National Laboratory, Argonne, IL 60439, USA*

P. Nevski, T. Wenaus
*Brookhaven National Laboratory, Upton, NY 11973, USA*

We present the design and the status of the database for detector description parameters in ATLAS experiment. The relational database for storing these "Primary Numbers" is based on the technology developed by some of the authors in the earlier project NOVA [1]. The ATLAS Primary Numbers are the parameters defining the detector geometry and digitization in simulations, as well as certain reconstruction parameters, including the identifier maps of the detector elements. Since the detailed ATLAS detector description needs more than 10,000 such parameters, a preferred solution is to have a single verified source for all these data. As a first approximation these parameters fit well into the structured name-value pair concept. A more detailed view of the parameter provides not only the parameter name and the value but also the parameter type (int, float,…), the textual comment describing the parameter meaning, the parameter scope (representing the structural container that this parameter belongs to), and may include additional information on the parameter unit of measurement and the parameter version. Effectively, the database stores the data dictionary for each parameter collection object, providing schema evolution support for object-based retrieval of parameters. The same Primary Numbers are served to many different clients accessing the database: the ATLAS software framework Athena, the Geant3 heritage framework Atlsim, the Geant4 developers' framework FADS/Goofy, the generator of XML output for detector description, and several end-user clients for interactive data navigation, including web-based browsers and ROOT. The preferred mode of access to these data is via connections to Athena services (conversion and conditions/transient stores). Another access mode is implemented through the interval-of-validity service for timestamp-based data retrieval within the conditions database. The choice of the MySQL database product for the implementation provides additional benefits: the Primary Numbers database can be used on the developers' laptop when disconnected (using the MySQL embedded server technology), with data being updated when the laptop is connected (using the MySQL database replication).

## 1. INTRODUCTION

### 1.1. LHC Computing Challenge

Beginning in 2007, the ATLAS experiment – a large general-purpose detector – will probe electroweak symmetry breaking and the origin of mass. The next generation of High-Energy Physics (HEP) experiments at the LHC presents new challenges to the field. Just the number of fast readout channels in these experiments is two orders of magnitude larger than in the previous generation of HEP experiments. Combined with the high data acquisition rates, the expected amount of raw detector data constitutes more than one petabyte per year for the ATLAS experiment alone. With reconstructed and simulated data the ATLAS total is about 10 PB/year – in one year ATLAS will produce more data than all of the HEP data that exists today. Compounding the problem is the very large, distributed nature of the collaboration – ATLAS is three times the size of the largest collaborations in running HEP experiments. A consequence of the unprecedented data sizes and the distributed nature of the collaboration and its computing infrastructure is the need for multiple advances in the tools for computing and analysis on unprecedented scales.

In addition to challenges in computing and data handling, the approaching generation of data-intensive experiments places a great burden on physicists and software professionals performing data processing and simulations and analysis to configure and manage the large number of parameters and options provided in the data processing software system. Management and processing of the primary physics event data is being largely addressed and made more transparent in the present round of LHC-wide grid computing projects. The ensuing progress in grid computing enables massive compute-intensive data processing capabilities. These powerful new resources need to be matched by equally powerful and intelligent user services that facilitate physicists' handling of numerous parameters that steer data processing and simulations.

### 1.2. Machine Data vs. "Human Data"

A key element in our approach to provide such services is the separation between two different kinds of data used in data-intensive sciences: the machine data and the "human data". The machine data – "the data" per se – are characterized by very large volumes, and by the fact that these data are acquired by scientific instruments or generated by computers. The "human data" are provided by scientists to control the data transformation of the machine data.

The principal focus of HEP computing has typically been upon the machine data. Being limited in volume, the "human data" were traditionally managed "by hand" and remained beyond the scope of the large grid computing projects. Because the "human data" must be thoroughly verified, their validation is a laborious iterative process similar to fundamental knowledge discovery.

The amount of "human data" in the LHC detector description domain makes LHC a challenging environment to test our approach for providing data management services for these data.

## 2. DETECTOR DESCRIPTION CHALLENGE

A state-of-the-art LHC detector (of the size of a five-storey building) is comprised of many elements. The detailed description of ATLAS detector geometry in simulations is unprecedented – the number of volumes in ATLAS detector simulations is an order of magnitude larger than in other experiments (Table 1).

Validation of parameters controlling the detector description is a time-consuming iterative process encapsulating considerable expert knowledge. To provide unprecedented detail in detector description, ATLAS requires more than ten thousand detector description parameters – clearly beyond the scope of manual data management capabilities. To provide a preferred solution for detector description we formalized the experiment's "institutional memory" in an ATLAS Primary Numbers database. In that approach, the Detector Description task in ATLAS is accomplished through the use of Primary Numbers that in simulations determine:

- detector geometry parameters,
- digitization parameters,

and in reconstruction define:

- the same detector geometry parameters,
- parameters of the reconstruction algorithms,
- alignment parameters.

Use of database technology ensures that the same Primary Numbers are served to many heterogeneous clients accessing the database. In addition, the navigability for the encapsulated knowledge is provided in interactive mode for the end-users.

Table 1: ATLAS Geometry Description Details

| Geometry Feature | Count |
|---|---|
| Volume Type | 4,673 |
| Volume Object | 23 K |
| Volume (Distinct Copies) | 29 M |

## 3. PRIMARY NUMBERS DATABASE

### 3.1. Primary Numbers

Innovative Virtual Data concepts from the GriPhyN project [2] introduced a generic aspect for many data-intensive science domains. Detector description in its several incarnations can be regarded as virtual data—data derived by transformations of the Primary Numbers that parameterize the ATLAS detector. Resulting detector description data in turn are integral to the definition of transformations of event data, such as those that produce hits from generator events, and digitizations from hits.

An important distinction between the "human data" and the machine data makes it clear that the Primary Numbers define the origin of all the massive machine data. In that regard the "human data" category is distinct from metadata and provenance, which are often simply machine data categories associated with data processing, historical information, and data dependencies tracking (though metadata may, in principle, also be used for annotation based upon human knowledge).

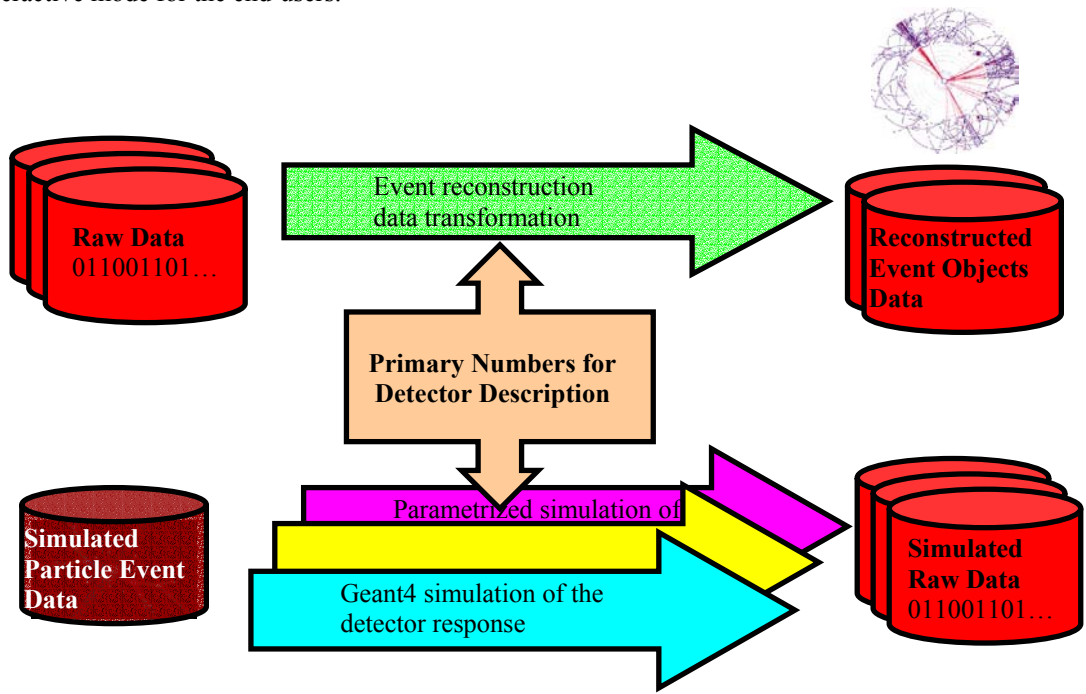

Figure 1. The Primary Numbers database plays a central role in ATLAS data processing architecture by providing unique detector description parameters for simulations and reconstruction data transformation.

**MOKT006**

Table 2: Example of Primary Numbers – Geant geometry parameters describing ATLAS Mother Volume

| Name | Value | Comment |
|---|---|---|
| Version | 2 | 2001 VERSION WITH ENDCAP SHIFTED B |
| Rmin | 0.0 | Inner Radius |
| Rmax | 1400.0 | Outer Radius |
| Zmax | 2350.0 | Maximum Z |

An often overlooked architectural aspect of the "human data" is their "primary" nature – these are the data that has to be provided to machines (detectors, computers) before any machine data comes out. In the ATLAS data processing architecture, the Primary Numbers database plays a central role by providing unique detector description parameters for simulations and reconstruction (Figure 1).

In a first approximation, the Detector Description parameters – the Primary Numbers providing control of applications simulating the detector response or performing the event reconstruction transformation – fit well into the XML concept of name-value pairs that can be arbitrarily structured to support content management (Table 2). A more detailed look at the Detector Description parameters reveals features extending beyond the simple name-value pairs concept. Facilitation of intelligent user services requires at least the following features:

- name of the parameter;
- value of the parameter;
- primitive type of the parameter (int, float,…);
- textual comment describing the parameter;
- scope of the parameter (representing the structural container that this parameter belongs to);

and may include additional information such as the parameter unit of measurement, the parameter version and the structural container information.

### 3.2. Encapsulating Validated Knowledge

Verification of the Primary Numbers requires considerable efforts: checking the engineering drawings, consulting with experts, etc. Our collaborator, Claire Bourdarios, reported that detailed analysis of ATLAS Liquid Argon calorimeter barrel accordion geometry parameters took about one month [3]. In total the barrel accordion Primary Numbers count was below one hundred. It took about two months to verify, extend and organize more than two hundred parameters capturing geometry of the largest pieces of ATLAS detector – toroid magnets and supports necessary to complete the muon system description [4].

Since the detailed ATLAS detector description needs more than ten thousand such parameters, a preferred solution is to establish a single verified source for all these data.

### 3.3. Schema Evolution

Due to the iterative nature of Detector Description validation and discovery, the content and the structure of the Primary Numbers database is necessarily evolving. To address the challenging problem of Detector Description evolution we choose relational database technology for the persistent data storage. In our relational approach, the internal data representation in the database is not a one-to-one copy of the transient data representation. In that regard this approach is distinct from some other methods that do not separate between the transient and persistent data representations.

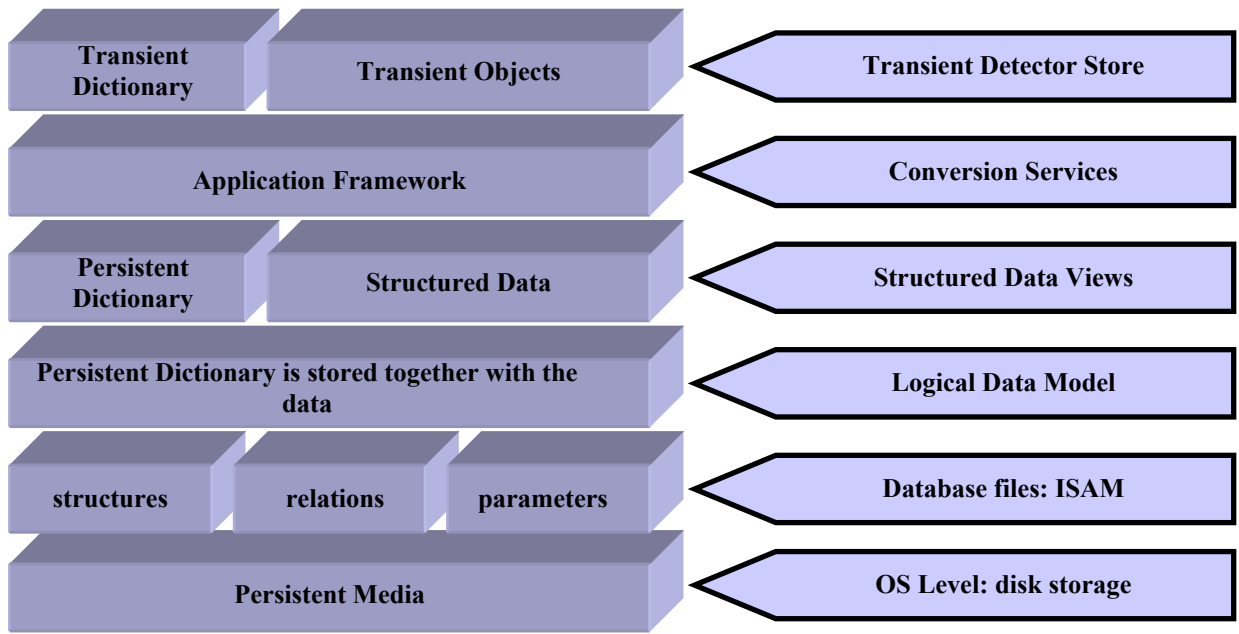

Figure 2. The multi-layered architecture provides separation between the transient and persistent objects.

**MOKT006**

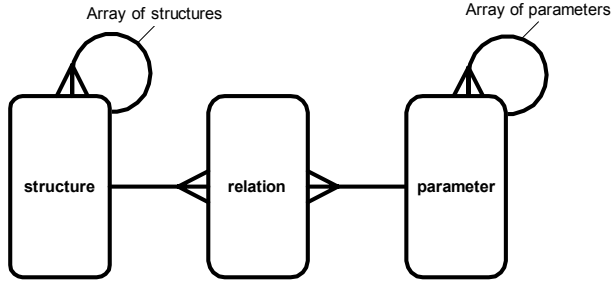

Figure 3: Data model for storing versioned objects in the relational database.

In the relational approach, when the structure of the transient data changes – the internal structure of the data in the relational database is not supposed to be changed. When such functionality is achieved, this is considered as a successful relational database design. Thus, the transient data structures are, in essence, the "views" of the persistent data representation. Effectively, the database stores the persistent data dictionary for each parameter collection object, providing schema evolution support for object-based retrieval of parameters. In our design this functionality is achieved through the multi-layered interface separating the transient and the persistent data representations (Figure 2).

### 3.4. Database Technology

For the Primary Numbers database we utilize the relational database technology that was introduced earlier in [1]. Figure 3 presents the data model used to provide generic data object storage.

Some Detector Description parameters are better handled in a binary format. Examples of binary data stored in the Primary Numbers database are the magnetic field maps and identifier maps of the detector elements used in simulations and reconstruction. Clearly, such data fall into the machine data category, as they result from computer calculations. To support the binary data storage we have extended the model of [1].

## 4. DATA ACCESS PATTERNS

### 4.1. Client Heterogeneity

The same Primary Numbers must be served to many heterogeneous clients accessing the database:
- ATLAS software framework Athena,
- Geant4 developers' framework FADS/Goofy,
- Geant3 heritage framework Atlsim,
- XML output generator for the detector description.

The encapsulated knowledge navigability is provided in interactive mode via the web browsers Netscape and Internet Explorer, as well as for ROOT clients.

Figure 4 presents key elements in our architecture to provide uniform data access for different clients. The main data access mode implemented in an earlier approach [1] where all the work was done in the object request broker base class has evolved to support a new grid services architecture, where service discovery requests and persistent-transient data conversions are now functionally separate. This new approach fits best within the Gaudi/Athena architecture separating the transient and persistent data representations.

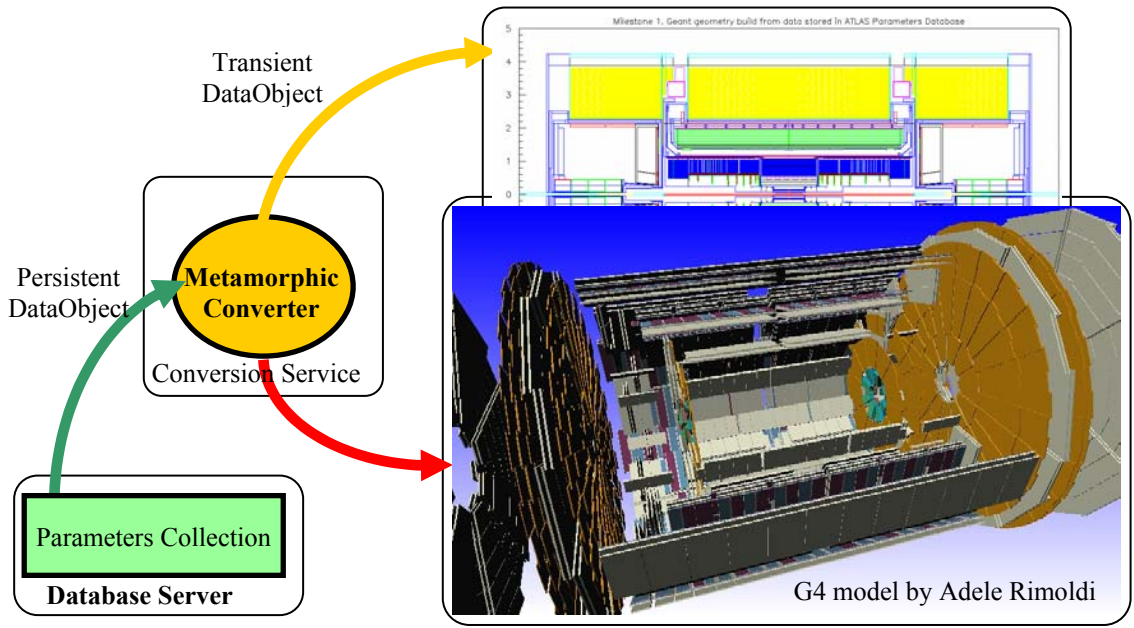

Figure 4. The key elements in the data access architecture providing uniformity for various clients.

## 4.2. Conversion Service

To deliver the new functionality we developed and deployed the NOVA Conversion Service providing access to Primary Numbers database within the ATLAS software framework Athena. The service registers Primary Number objects with the transient detector store. The objects then can be accessed in the usual StoreGate manner (via retrieve).

The conversion service is supported by automatic Primary Numbers database update procedures through the generic data object dictionary interface and utilities for generation of headers and converters for more than two hundred classes enhanced with reflection capabilities to support Primary Number objects encapsulating evolving detector description knowledge.

## 4.3. IOV/ConditionsDB Interface

In the ATLAS software framework Athena, the high-level view for non-event data is embodied in the Transient Detector Store concept. In this architecture all of the non-event data, e.g., calibrations and alignment data are accessed through the uniform Conditions Database interface supporting automatic data retrieval based upon their interval-of-validity (IOV) in relation to the timestamp of the corresponding event data.

Together with our collaborators, RD Schaffer and Antoine Pérus, we demonstrated that the IOV database could provide validity-interval-based access to data that has been stored using an existing NOVA Conversion Service. Figure 5 presents the essential elements of our solution.

The essential functionalities delivering Primary Numbers Services integration behind the IOV/ConditionsDB Interface are comprised of the following:
- the IOV database associates a folder name (e.g., "Tile/Pedestals"), an interval of validity [ti, tj), and a tag to a String;
- the String will contain an externalized IOpaqueAddress;
- IOV integration with Primary Numbers database requires that NOVA services:
  - be capable of storing multiple object instances of the same type,
  - assign unique IOpaqueAddresses (instance names),
  - externalize IOpaqueAddresses as strings, and "internalize" strings to create IOpaqueAddresses.

These capabilities would ideally be added to base classes for all conversion services.

An important benefit of this approach is that clients reading data do not need to know which conversion service (e.g., NOVA) is delivering the data. The StoreGate retrieve will typically use a key that is the folder name (e.g., "Tile/Pedestals"); the timestamp information will come from the current event without user intervention (transient IOV service). Internally, this will be enough (along with a tag set in jobOptions) to retrieve the correct string; the string will be changed to an IOpaqueAddress; the usual Athena/Gaudi conversion service mechanisms will be triggered to build an object in the transient store — all of this is hidden from the user.

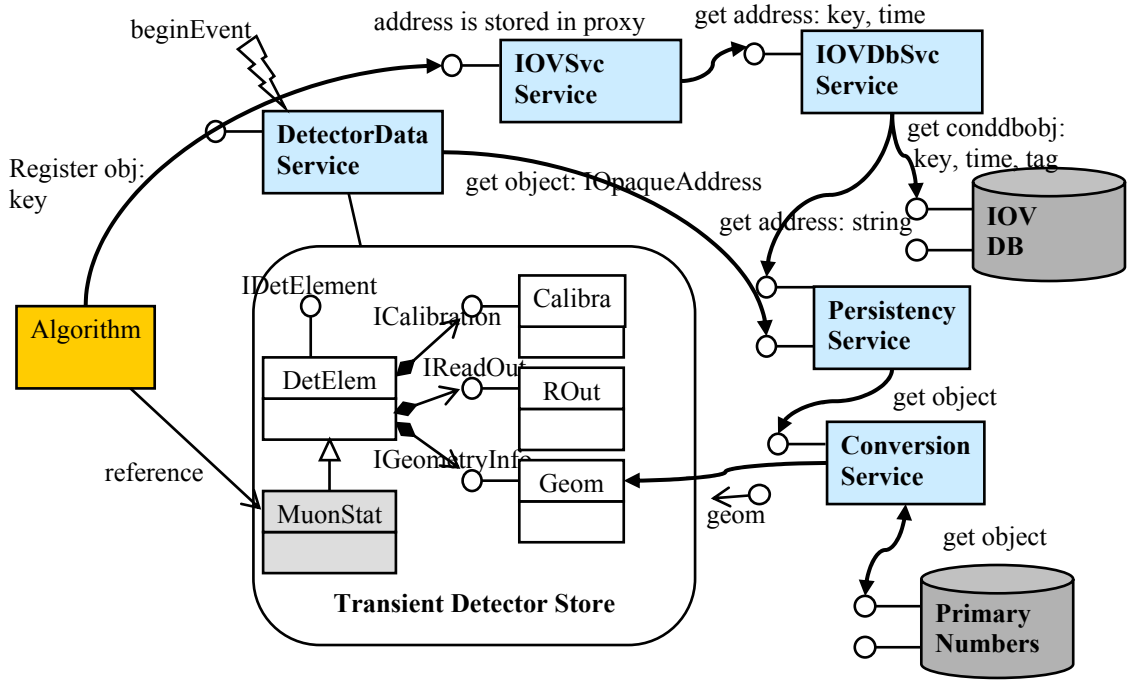

Figure 5. Integrated solution for the Detector Store updates using IOV database interface to the Primary Numbers database.

**MOKT006**

The intent of the provided solution is to do nothing that will break or change current NOVA services use that does not require intervals of validity: the IOV database provides additional capabilities, but does not replace current mode of access to data in the Primary Numbers database.

## 5. MYSQL IMPLEMENTATION BENEFITS

The choice of the MySQL database product for our initial implementation added several specific benefits. The Primary Numbers database can be used on a developer's laptop when disconnected (using the MySQL embedded server technology), with data being updated when the laptop is connected (using the MySQL database replication capabilities). Another useful feature is the improved support for binary data transfers provided in the newer version of MySQL server software. We have also tested MySQL certificate authorization technologies, which we have found useful in the emerging computing grid environment.

The absence of Open Grid Service Architecture (OGSA) compatible database services in the first round of grid testbed deployment calls for interim solutions to the database access problems exhibited on many grid sites. We have found MySQL functionality very useful in resolving such problems. The preferred solution is the use of embedded server technology for applications running in a closed network environment. Another alternative emphasizing uniformity of applications for all environments is the use of the extract-transport-install components of the ATLAS database architecture to deliver database services to the Compute Elements behind closed firewalls via the available grid transport mechanisms. In collaboration with ATLAS grid software developers we have successfully tested such a solution to enable ATLAS Data Challenge 1 production on the NorduGrid computing facilities [5].

## 6. ROADMAP TO SUCCESS

An important new capability of the Primary Number database will be the integration of its services into the future framework of OGSA-compatible database services. Our persistency solutions for both the domain of Primary Numbers for Detector Description and the domain of production recipe catalogs (Virtual Data Cookbook) for data management [6] encapsulate valuable expert knowledge acquired in a time-consuming iterative process similar to the fundamental knowledge discovery. We envision that coherent database solutions formalizing extensible "institutional memory" will provide a foundation for future scalable Knowledge Management Services that will innovatively integrate two advancements in Computing Sciences – Grid Computing technology, providing access to vast resources, and novel meta-computing approaches from the Software Assurance community, to deliver knowledge navigability, accessibility and assurance to the data management framework.

## Acknowledgments

We thank all of our ATLAS collaborators and, in particular, all the developers from the Simulation and Detector Description domains whose participation was instrumental to enable rapid prototyping cycles and early users' feedback. We thank our database collaborators RD Schaffer and Antoine Pérus for their contribution in establishing IOV services. We also wish to thank Claire Bourdarios, Andrea Dell'Acqua, Michael Lelchouk, Armin Nairz, Lawrence Price and Adele Rimoldi for numerous discussions, interest and support of this work. The Argonne National Laboratory's work was supported by the U.S. Department of Energy, Office of Science, Office of High Energy and Nuclear Physics, and Office of Advanced Scientific Computing Research, under U.S. Department of Energy contract W-31-109-Eng-38.